\documentclass[graybox]{svmult}

\usepackage{type1cm}        
\usepackage{makeidx}         
\usepackage{graphicx}        
\usepackage{multicol}        
\usepackage[bottom]{footmisc}

\usepackage{newtxtext}       %
\usepackage[varvw]{newtxmath}     

\makeindex             

\usepackage{xcolor}
\usepackage{url} 
\usepackage{hyperref} 
\hypersetup{colorlinks,linkcolor={blue},citecolor={blue},urlcolor={blue}} 

\begin{document}

\title*{Faster rate of Hawking radiation in modified gravity constraining dark matter}
\titlerunning{The Relativistic Universe: From Classical to Quantum}
\author{Panchajanya Dey and Banibrata Mukhopadhyay}
\authorrunning{Dey \& Mukhopadhyay}
\institute{Panchajanya Dey \at Department of Physics, Indian Institute of Science, Bangalore, \email{panchajanyad@iisc.ac.in}
\and Banibrata Mukhopadhyay \at Department of Physics, Indian Institute of Science, Bangalore, \email{bm@iisc.ac.in}}
%
%
\maketitle
\textit{To be published in 
Astrophysics and Space Science Proceedings, titled "The Relativistic Universe: From Classical to Quantum, Proceedings of the International Symposium on Recent Developments in Relativistic Astrophysics", Gangtok, December 11-13, 2023: to felicitate Prof. Banibrata Mukhopadhyay on his 50th Birth Anniversary", Editors: S Ghosh \& A R Rao, Springer Nature}

\vskip1.0cm
\abstract{
	The exact theory of gravity in the strong field regime is still under debate. There are 
	observations implying the need for modification to Einstein's gravity.
	On the other hand, the exact constituents of dark matter are also a big puzzle, 
 where primordial black holes (PBHs) are argued to be a potential candidate. We explore Hawking radiation in a modified gravity and find that PBHs evaporate faster in a scalar-tensor theory based modified gravity. Subsequently, all the nonrotating BHs
	of mass $\sim 10^{15}$ g or less should have been evaporated by today, which is an order of magnitude heavier
	than what the Einstein gravity predicts. This has many consequences including a strict constraint 
	on contributing PBHs to dark matter, widening the debate of dark matter origin.
}



\section{Introduction}


The most popular theory of cosmology is the $\Lambda$-CDM model when the 
universe contains $5\%$ ordinary matter, $26.8\%$ dark matter, and $68.2\%$ dark energy.
The dark matter is assumed to be cold (of low velocity), which exhibits structures of the universe emerging by the gradual 
accumulation of particles. Nevertheless, there has been no success in the discovery of dark matter particles even 
over the half a century search. This and more recent observations lying with gravitational wave and James Webb 
Space Telescope have considerably strengthened the idea of primordial and direct collapse black holes (BHs)
to be dark matter \cite{bird,jwst}.
Moreover, intrigued by specific observations not explainable properly by ordinary dark matter (as well as
dark energy) hypothesis, 
various modifications to Einstein's theory of general relativity (GR), including modified Newtonian dynamics 
(MOND) \cite{mond} were proposed. 
However, no modified gravity (MG) theory seems to describe every piece of observation successfully 
at the same time. This suggests that even in the presence of MG, the existence of dark matter is 
inevitable.

GR is the most panoramic theory
that explains the theory of gravity. It can easily explain a large number of phenomena where
Newtonian gravity falls short, which includes
deflection of light in strong gravity, generation of gravitational wave, gravitational redshift
of light etc. However, apart from dark matter, a number of other recent observations in cosmology and
in the high density/energy regions in the universe, such as the inside/vicinity of compact objects, it seems
that the GR may not be the ultimate theory of gravity \cite{psaltis,ups}. Starobinsky
first used one of the modified theories of gravity, namely $R^2$-gravity \cite{staro}, to 
explain dark energy era and the very early universe cosmology, thereby did
overcome some shortcomings of GR to explain cosmos. Eventually, plenty of different models have been
proposed to explain various physical phenomena where GR
fails. The question, hence, arises whether globally GR is the ultimate theory of
gravitation, or it requires modification in the strong gravity regime. 
In stellar physics, indeed it was shown that MG theories reveal significant deviations from
the GR solutions of neutron stars (NSs) \cite{dam,eksi,aneta}.
However, MG was also explored in BHs, e.g. asymptotically flat vacuum solutions \cite{kalitaMukho}. 
The present work centers around MG based BHs.

When a MG is introduced, whatever be the (observational) purposes, it should suffice other experiments/observations,
which do not necessarily need a MG. 
That is, it should pass through the solar system test or at some regime
any MG should reduce to GR (and subsequently to special relativity theory). In other words, if MG is true for 
one physics and astrophysics, it should be true for other gravity involved physics as well. Some of the interesting
gravity induced sites are, NS interior, early evolution of universe, Hawking radiation, BH accretion flow,
particle creation in curved spacetime, gravitational lensing and waves, to say a few. 

One of the very interesting, old and phenomenal physical processes lying with BHs is the Hawking radiation.
All the BHs keep radiating continuously and, hence, get evaporated. The rate of evaporation is 
inversely proportional to the mass of BHs. Hence, the primordial BHs (PBHs) being much lighter than
the ones formed by the stellar collapse evaporate faster. 
According to GR, all the BHs of mass $\lesssim 5\times 10^{14}$ g forming in early universe have been
evaporated by today. On the other hand, a part of PBHs is argued to be the candidate for dark matter \cite{rajeev}.

What if the gravity is MG not GR? What is the fate of Hawking radiation then, evaporation of PBHs
and its connection to dark matter? Can it resolve the problem of dark matter origin? Can it enlighten the issues of the invisible source of gravity in the universe? We need to explore GR/MG, BH physics, cosmology, and a bit of particle physics in a unified manner to answer this question.\\

\section{Brief Description of a modified gravity metric}

There are several ways to introduce modification to GR. One of such ideas is based on scalar-tensor
theory \cite{sca-ten} which further leads to a special class of MG called $f(R)$-gravity. 
With the metric signature $(+,-,-,-)$, the modified Einstein-Hilbert action is given by
\begin{equation}
    S = \int \left[\frac{c^4}{16\pi G}f(R)+\mathcal{L}_M\right]\sqrt{-g}d^4 x,
\end{equation}
where $f(R)$ is a function of $R$, which is Ricci scalar, $-g$ is the determinant of metric tensor $g_{\mu\nu}$, ${\cal L}_M$ the matter Lagrangian density, $G$ Newton's gravitation constant, and $c$ the 
speed of light.
This gives MG equation as
\begin{equation}
    F(R)G_{\mu\nu}+\frac{1}{2}g_{\mu\nu}[RF(R)-f(R)]-(\nabla_\mu\nabla_\nu-g_{\mu\nu}\Box)F(R) = \frac{8\pi G}{c^4}T_{\mu\nu},
\end{equation}
where $F(R)=df(R)/dR$, $G_{\mu\nu}$ is the Einstein tensor, and $T_{\mu\nu}$ the energy-momentum tensor. For $f(R)=R$, MG reduces to GR. Assuming a spherically symmetric vacuum solution of MG equation given by
\begin{equation}
    g_{\mu\nu} = diag\biggl(s(r),-p(r),-r^2,-r^2sin^2\theta\biggr)
\end{equation}
and also assuming that $F(R) = 1+ B/r$ \cite{kalitaMukho}, such that the vacuum solution resembles GR 
at large distance, i.e. $r\rightarrow\infty$, 
we obtain
\begin{eqnarray}\nonumber
	s(r) &=& 1 - \frac{2}{r} -\frac{B(-6 + B)}{2r^2} + \frac{B^2(-66 + 13B)}{20r^3} - \frac{B^3(-156 + 31B)}{48r^4} + \frac{3B^4(-57 + 11B)}{56r^5} \\
	&-& \frac{B^5(-360 + 67B)}{128r^6} + \mathcal{O}(r^{-7})
	\label{gtt}
\end{eqnarray}
and $p(r) = X(r)/s(r)$, where $X(r)$ is given by
\begin{eqnarray}
    X(r) = \frac{16r^4}{(B+2r)^4},
	\label{grr}
\end{eqnarray}
for $B<0$. Note that $B>0$ corresponds to $R<0$ which implies unphysical repulsive gravitational field, hence is ruled
out. Equations (\ref{gtt}) and (\ref{grr}) reduce to those of GR (Schwarzschild metric) either for $B=0$
or at $r\rightarrow\infty$.\\

\section{Size of black hole in modified gravity}

GR suggests the radius of a nonrotating BH to be $2GM/c^2$, which is called Schwarzschild radius, also is event horizon ($r_H$). To obtain the event horizon radius in MG, we equate the denominator of \(g_{rr}\), i.e. $-p(r)$, to zero. 
For the present purpose, the event horizon corresponds to $s(r) = 0$, leading to $r_H$ in MG upto the first order 
in $B$ as
\begin{equation}
    r_H = 2-1.5B+\mathcal{O}(B^2).
\end{equation}
In general, the dependence of $r_H$ on the parameter $B$ is shown in Fig. 1.
It is seen therein that as $B$ decreases, $r_H$ increases. This
implies that MG increases the effective mass of BH, if compared with a Schwarzschild BH of the same $r_H$.\\

\begin{figure}
    \centering
    \includegraphics[height = 9 cm]{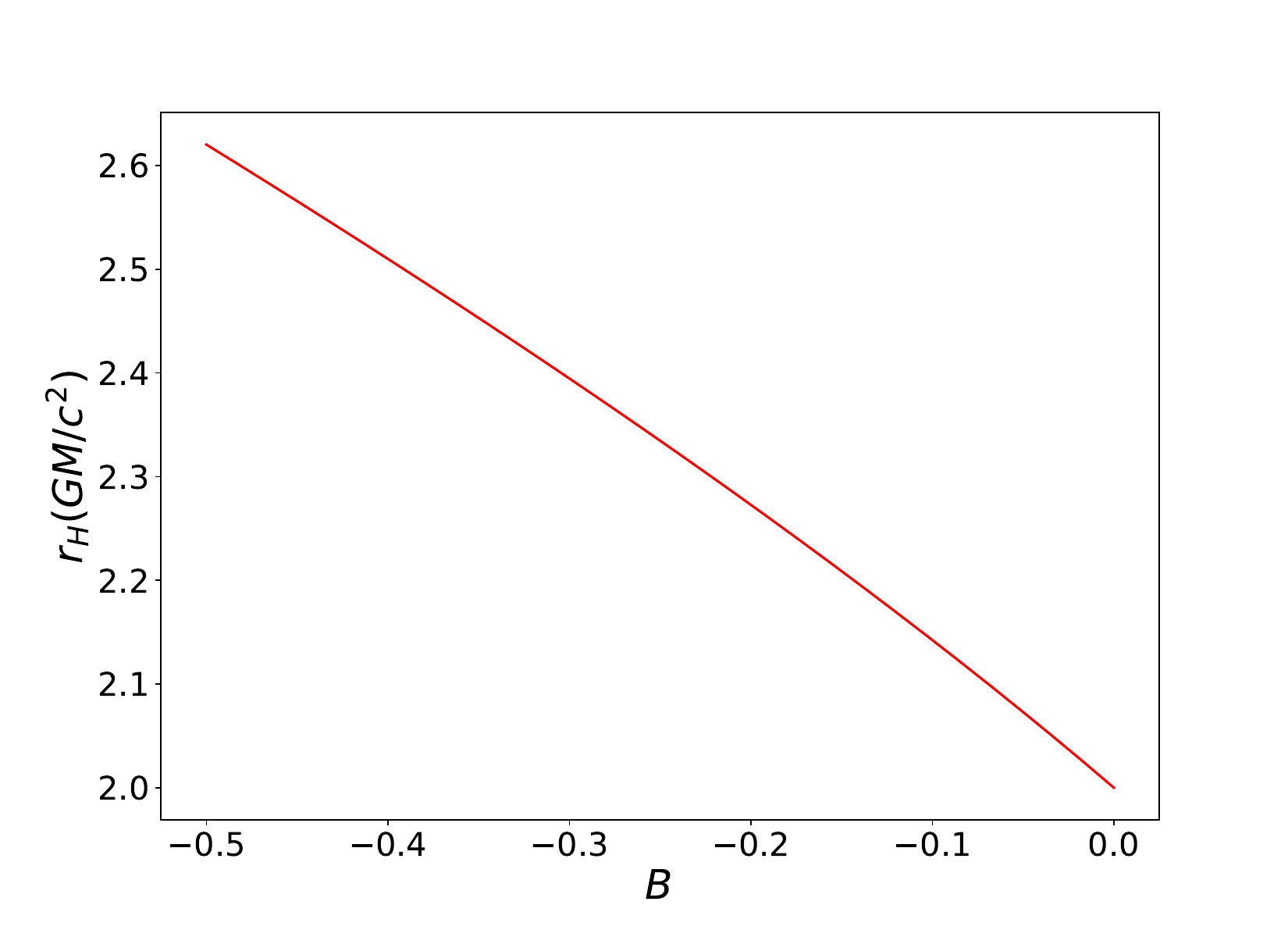}
    \caption{Radius of event horizon of a nonrotating BH as a function of $B$.}
    \label{fig1}
\end{figure}

\section{Hawking temperature in a modified gravity}

The Hawking temperature of a BH in GR can be defined as
%
\begin{equation}
    T_H = \frac{\hbar\kappa}{2\pi ck_B},
\end{equation}
where $\kappa$ is the surface gravity, which in GR for a nonrotating BH is given by $\ddot{r}=\kappa=c^4/4GM$.
In general, 
\begin{equation}
    \kappa^2 = \frac{1}{4}\frac{s(r)'^2}{s(r)p(r)},
\end{equation}
with $c=k_B=1$, which in our MG leads to (for  $B \ll 0$)
\begin{equation}
    \kappa = \frac{1}{r_H^2}\left(1-\frac{2B}{r_H}\right).
\end{equation}
\begin{figure}
    \centering
    \includegraphics[height=9 cm]{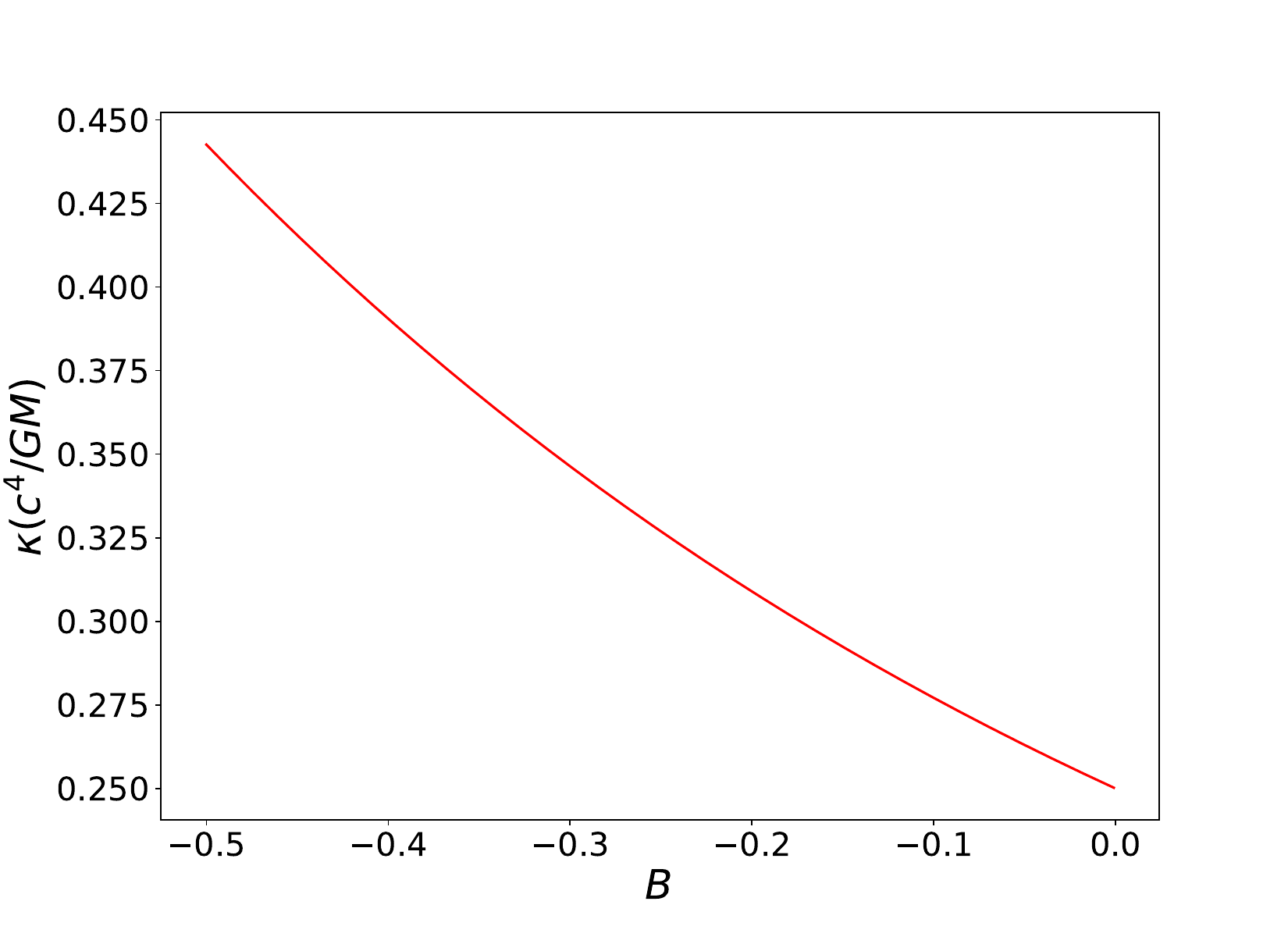}
    \caption{Surface gravity as a function of $B$ for a nonrotating BH.}
    \label{fig2}
\end{figure}
Therefore, while in GR, $r_H$ and $\kappa$ decrease with the increase of specific angular momentum and charge of the BH for a given mass, in MG, both of them increase with the decrease in $B$ (i.e. magnitude of $B$ increases which implies stronger modifications to the gravity), see Figs. 1 and 2. This implies that the temperature of the BH in MG increases. 

Now we assume the BH to be a grey-body having the temperature and surface area respectively equal to the Hawking temperature and event horizon area of the BH. 
If the grey-body parameter is denoted by $\Gamma$, in GR the rate of evaporation of a BH is
\begin{equation}
    Q_{GR} = -c^2\frac{dM}{dt} = 4\pi\Gamma r_{H}^2 \sigma T_H^4,
    \label{eq8}
\end{equation}
where $\sigma$ is the Stefan-Boltzmann constant.
Thus, in GR the time taken for a BH of mass $M$ to evaporate completely is
\begin{equation}
	t_{GR} = 2.1 \times 10^{67} {\rm yrs } \times\frac{1}{\Gamma} \left(\frac{M}{M_\odot}\right)^3.
    \label{eq9}
\end{equation}
Now in MG, the temperature and radius change. Therefore, we define the ratio of the temperatures in MG to GR as $\Theta = T_{MG}/T_{GR}$, and the corresponding radii ratio as $\rho = r_{MG}/r_{GR}$. Therefore, the rate of evaporation of a BH in MG can be written in terms of the same rate in GR as
\begin{equation}
    \frac{Q_{MG}}{Q_{GR}} = \xi = \rho^2\Theta^4.
\end{equation}
The variation of $\xi$ with $B$ is shown in Fig. 3.
\begin{figure}
    \centering
    \includegraphics[height = 9 cm]{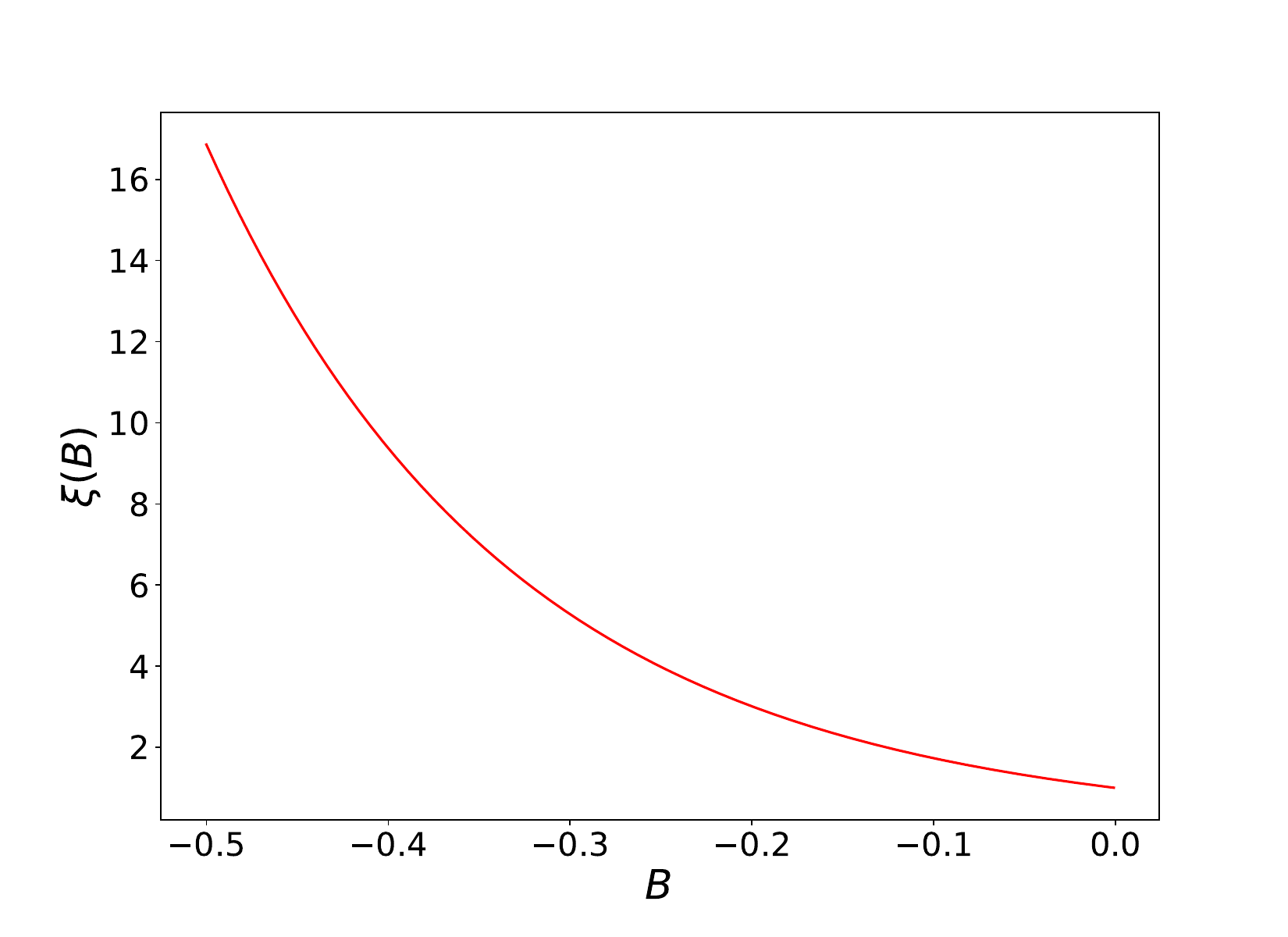}
    \caption{The variation of $\xi$ as a function of $B$.}
    \label{fig3}
\end{figure}
Therefore, the time of evaporation ($t_{MG}$) changes in MG w.r.to that in GR as
\begin{equation}
	\frac{t_{MG}}{t_{GR}} = \frac{1}{\xi} = \frac{1}{\rho^2\Theta^4},
    \label{eq11}
\end{equation}
which leads to the evaporation time of a BH of initial mass $M$ in MG as
\begin{equation}
	t_{MG} = 2.1 \times 10^{67} {\rm yrs } \times\frac{1}{\Gamma\rho^2\Theta^4} \left(\frac{M}{M_\odot}\right)^3.
\end{equation}
Since, both $\rho$ and $\Theta$ are greater than unity, the time of evaporation of a nonrotating BH clearly decreases in MG compared to that in GR.\\

\section{Minimum survival mass of primordial black hole}

In GR, the maximum initial mass of a PBH that evaporates completely in the current age of the universe can be computed easily based on the above discussion and grey-body assumption, which is $2.5\times10^{14}$ g, or $1.25\times10^{-19} M_\odot$ \cite{Carr_2021}. In MG, since the rate of evaporation is more than that of GR by a factor of $\xi$, the mass of PBH which will be evaporated by today is $\xi\times 2.5\times10^{14}$ g, according to the chosen MG theory. 
The dependence of the maximum evaporated mass of PBH on $B$ is shown in Fig. 4.
\begin{figure}
    \centering
    \includegraphics[height = 9 cm]{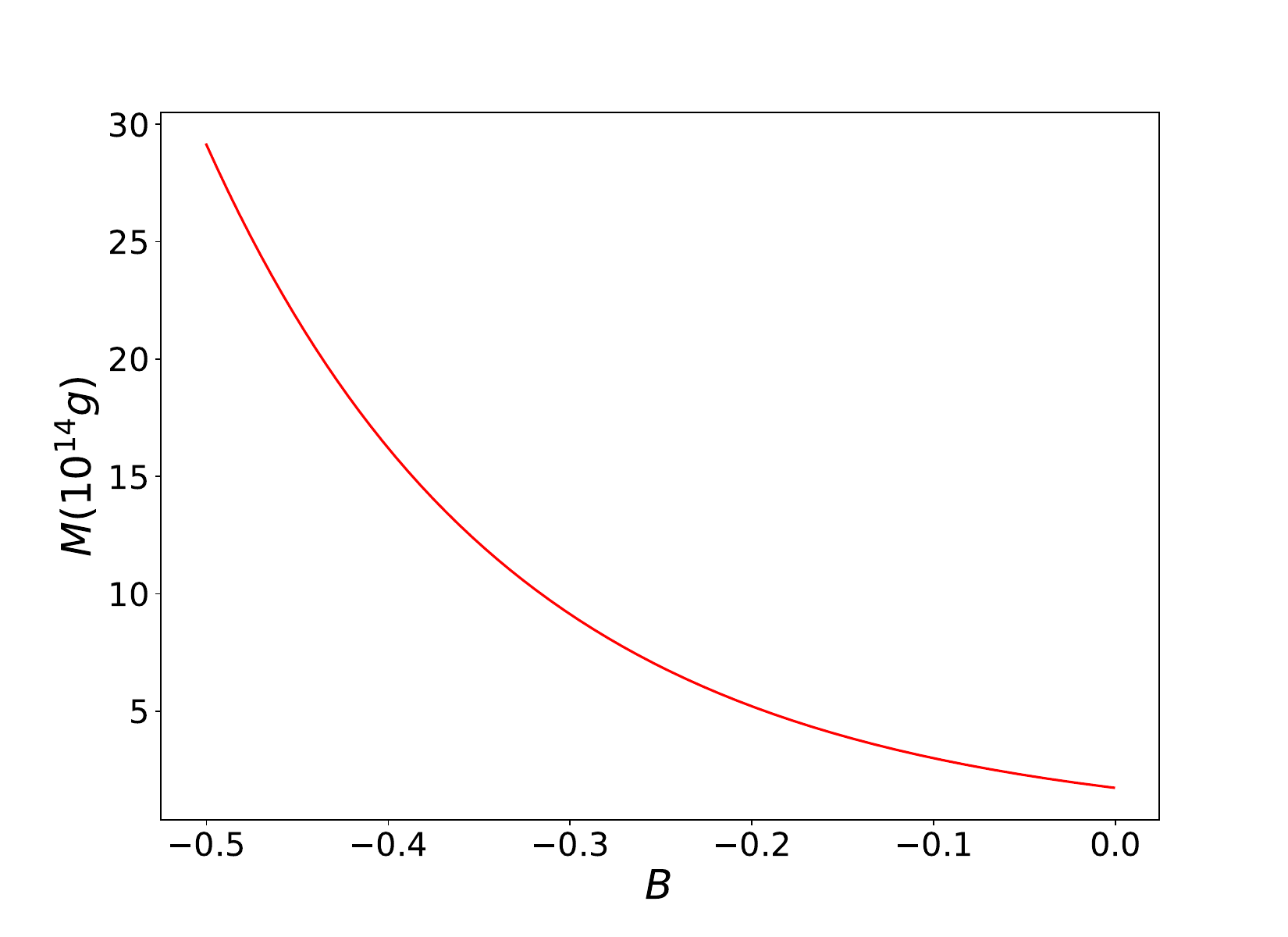}
    \caption{Maximum initial PBH mass that completely evaporated today as a function of $B$.}
    \label{fig4}
\end{figure}
Depending on $B$, the mass of PBH that is expected to be completely evaporated by today can be even an order of magnitude higher than that predicted by GR, even in the given restricted range of $B$.\\

\section{Implications of the result}

In GR, PBHs which have a mass much larger than $10^{15}$ g are practically unaffected by Hawking radiation over the age of the universe. Hence, PBHs of mass $\gtrsim 10^{15}$ g are considered to be a potential dark matter candidate which makes upto $26.8\%$ of the critical density. On the other hand, PBHs would have formed much earlier than the end of the radiation-dominated era. Therefore, they would not be affected by the well-known BigBang Nucleosynthesis (BBN) constraint that the baryons can have at most 5\% of the critical density, hence PBH contribution to critical density and dark matter would be non-baryonic. As Hawking radiation affects the survival of PBHs in MG, it plays an important role in constraining the contribution of PBHs to the dark matter, because massive PBHs are more constrained to
contribute to dark matter \cite{Carr_2021}. For $B = -0.47$, the minimum mass of PBH that survives evaporation through Hawking radiation increases $10$ folds, indicating a strong dependence of the minimal survival mass on the MG parameter $B$.\\


According to MG, a PBH can survive a lesser time. Therefore, we can expect that the faster evaporation in MG constrains on PBH as a potential candidate for the dark matter more strictly. This offers the scope of alternative theories, including axions and Weakly Interacting Massive Particles (WIMPs), to be better in a MG model, even though they are not yet discovered. Hence, the present work further widens the debate on the origin of dark matter. Although in these calculations, we have assumed that the age of the universe is independent of the gravity theory, this needs to be verified. Further, the rotating BHs would add up constraints, what needs to be explored in future.

\end{document}